\documentclass[
amsmath,
amssymb,
prl,
twocolumn,
showpacs,
superscriptaddress,
preprintnumbers]{revtex4}
\usepackage{mathrsfs}

\newcommand{\Lagrangian}{\mathscr{L}}

\newcommand{\beq}{\begin{equation}}
\newcommand{\eeq}{\end{equation}}
\newcommand{\bea}{\begin{eqnarray}}
\newcommand{\eea}{\end{eqnarray}}
 
\def\<{\langle}
\def\>{\rangle}
\def\d{\partial}
\def\+{\dagger}

\newcommand\A{{\mathrm{A}}}
\newcommand\f{{\mathrm{f}}}
\newcommand\B{{\mathrm{B}}}
\renewcommand\L{{\mathrm{L}}}
\newcommand\R{{\mathrm{R}}}

\begin{document}

\preprint{INT-PUB 04-13}
\title{Quantum Anomalies in Dense Matter}
\affiliation{Institute for Nuclear Theory, University of Washington,
Seattle, Washington 98195-1550}
\affiliation{Department of Physics and Astronomy,
University of British Columbia,
Vancouver, BC, Canada, V6T 1Z1}
\author{D.\,T.~Son}
\email{son@phys.washington.edu}
\affiliation{Institute for Nuclear Theory, University of Washington,
Seattle, Washington 98195-1550}
\author{Ariel~R.~Zhitnitsky}
\email{arz@physics.ubc.ca}
\affiliation{Department of Physics and Astronomy,
University of British Columbia,
Vancouver, BC, Canada, V6T 1Z1}
\date{May 2004}
\begin{abstract}
We consider the effects of quantum anomalies involving the baryon
current for high-density matter.  In the effective Lagrangian, the
anomaly terms describe the interaction of three light fields: the
electromagnetic photons $A_{\mu}$, neutral light Nambu-Goldstone
bosons ($\pi$, $\eta$, $\eta'$), and the superfluid phonon. The
anomaly induced interactions lead to a number of interesting phenomena
which may have phenomenological consequences observable in neutron
stars.
\end{abstract}
\maketitle

{\em Introduction}.---%
The rich phase diagram of QCD at high baryon density has attracted
considerable attention recently~\cite{Rajagopal:2000wf}.  Much of
attention has been paid to the determination of possible symmetry
breaking patterns, which are important for understanding the
low-energy dynamics.  A typical example is QCD with three light
flavors.  The ground state [the color-flavor-locked (CFL) phase] was
determined to break baryon number, chiral and U(1)$_\A$ symmetries.
One then writes down an effective Lagrangian and find it parameters by
matching calculations.

On the other hand, effects originating from quantum anomalies have not
been systematically studied (exceptions include
Refs.~\cite{Nowak:2000wa,Forbes:2001gj}).  It is well known that
anomalies have important implications for low-energy physics: the
electromagnetic decay of neutral pions is a textbook example.  In this
Letter we investigate the roles of anomalies at finite density.  In
contrast to previous works, we concentrate on the anomalies involving
the baryon current.  The effects coming from these anomalies are
strikingly unusual; they reveal intricate interactions between the
topological objects such as vortices and domain walls, the
Nambu-Goldstone (NG) bosons and gauge fields.  We mention here only
two effects: (i) there are classical weak neutral currents flowing on
superfluid vortices; (ii) there are electric currents flowing on
U(1)$_\A$ vortices, which exist at high energies~\cite{SSZ-domain}.
More effects are considered in the paper.

  
{\em Anomalies}.---%
One method to derive anomalous terms in effective theories is to put
the system in external background gauge fields and require that the
effective theory reproduces the anomalies of the microscopic theory.
We thus consider QCD in the background of two U(1) fields: the
electromagnetic field $A_\mu$ and a fictitious (spurion) $B_{\mu}$
field which couples to the baryon current.  Only at the end of the
calculations we will put $B_\mu=\mu n_\mu$, $n_\mu=(1, \vec{0})$,
corresponding to a finite baryon chemical potential.  (The technique
resembles the one used in Refs.~\cite{Kogut:1999iv,Kogut:2000ek} in a
similar context.)  The Lagrangian describing the coupling of quarks
with $B_\nu$ is
\begin{equation}
\label{1}
  \Lagrangian_\B = \bar \psi \gamma^\nu \left(i\d_\nu
    +{\textstyle \frac13} B_\nu\right) \psi
\end{equation}
(the baryon charge of a quark is taken to be $1/3$).  The Lagrangian
is invariant under U(1)$_\B$ gauge transformations
\begin{subequations}\label{qBU1B}
  \begin{align}
    q &\rightarrow e^{i\beta/3} q, \label{qU1B} \\
     B_{\mu} &\rightarrow B_{\nu}+\partial_{\mu}\beta,
  \end{align}
\end{subequations}
The QCD Lagrangian is also invariant under Lorentz transformations,
assuming $B_\nu$ transforms as a vector.

Let us assume that the baryon number symmetry is spontaneously broken.
The low-energy dynamics then contains a NG boson $\varphi_\B$, which
is the U(1)$_\B$ phase of the condensate.  This mode is the superfluid
phonon and exists in both the CFL phase and in nuclear matter with
nucleon pairing.  The transformation property of $\varphi_\B$ can be
found from Eq.~(\ref{qU1B}),
\begin{equation}\label{phiU1B}
  \varphi_\B\to\varphi_\B + 2\beta
\end{equation}
where the factor 2 is the baryon charge of the U(1)$_\B$ breaking
order parameter (assumed to have the baryon charge of a Cooper pair of
two baryons).  We also assume the there exist a neutral NG boson which
comes from breaking of a chiral symmetry.  This can be a $\pi^0$,
$\eta$ or $\eta'$ boson.  To keep our discussion general, we introduce
the current which creates this boson,
\begin{equation}\label{JA}
  j^\A_\mu = \frac12
  \sum_{i=1}^{N_\f} Q^5_i\, \bar\psi_i\gamma_\mu\gamma^5\psi_i
\end{equation}
For example, for $\pi^0$, $Q^5_u=-Q^5_d=1$, $Q^5_s=0$, and for
$\eta'$, $Q^5_u=Q^5_d=Q^5_s=1$. This current generates a chiral
transformation,
\begin{equation}\label{qU1A}
  \psi_i \to\exp\Bigl(\frac i2 \alpha Q^5_i\gamma^5\Bigr)\psi_i\,.
\end{equation}
The NG boson is characterized by a phase $\varphi_\A$, which
transforms under~(\ref{qU1A}) as
\begin{equation}\label{phiU1A}
  \varphi_\A \to \varphi_\A + q_\A\alpha\,,
\end{equation} 
where $q_\A$ is a number characterizing the axial charge of the
condensate.  The factor $\frac12$ was introduced in Eqs.~(\ref{JA})
and (\ref{qU1A}) so that the U(1)$_\A$ charge of the chiral condensate
$\<q_\L\bar q_\R\>$ is one.  In color superconducting phases, the
current will be normalized so that $q_\A=2$.

The effective low-energy description must respect the U(1)$_\B$ gauge
symmetry and be invariant under the gauge transformations~(\ref{qBU1B})
and (\ref{phiU1B}).  Therefore, in the effective Lagrangian the
derivative $\d_\mu\varphi_\B$ should always appear in conjunction with
$B_\mu$ to make a covariant derivative $D_\mu\varphi_\B =
\d_\mu\varphi_\B-2B_\mu$.  Vice versa, each occurrence of $B_\mu$ has
to accompanied by a $-\frac12\d_\mu\varphi_\B$.  The effective
Lagrangian should be relativistically invariant before the replacement
$B_\mu\to(\mu,\vec{0})$ is made~\cite{Son-superfluid}.

In the absence of background fields, neglecting the quark masses, the
chiral current~(\ref{JA}) is conserved if $\sum_i Q^5_i=0$.  The
conservation is violated for $\sum_i Q^5_i\neq0$ by instanton effects,
but at sufficiently high density such effects are suppressed.  In the
presence of the background electromagnetic and U(1)$_\B$ fields, the
conservation of (\ref{JA}) is violated by triangle anomalies:
\begin{equation}\label{QCD-anom}
\begin{split}
  \d^\mu j^\A_\mu &= -\frac1{16\pi^2}\bigl(
  e^2 C_{\A\gamma\gamma} F^{\mu\nu}\tilde F_{\mu\nu}
  - 2 e C_{\A\B\gamma} B^{\mu\nu}\tilde F_{\mu\nu}\\
  &\qquad+ C_{\A\B\B} B^{\mu\nu}\tilde B_{\mu\nu}\bigr)
\end{split}
\end{equation}
where $F_{\mu\nu}=\d_\mu A_\nu-\d_\nu A_\mu$ and $B_{\mu\nu}=\d_\mu
B_\nu-\d_\nu B_\mu$; $\tilde F_{\mu\nu} =
\frac12\epsilon_{\mu\nu\alpha\beta}F^{\alpha\beta}$, $\tilde
B_{\mu\nu} = \frac12\epsilon_{\mu\nu\alpha\beta}B^{\alpha\beta}$.  The
coefficients $C$'s in Eq.~(\ref{QCD-anom}) are given by
\begin{equation}\label{Cs}
\begin{split}
  & C_{\A\gamma\gamma} = 3\sum\nolimits_i Q^5_i (Q_i)^2,\quad
  C_{\A\B\gamma} =  \sum\nolimits_i Q^5_i Q_i,\\
  & \qquad\qquad C_{\A\B\B} = \frac13\sum\nolimits_i Q^5_i
\end{split}
\end{equation}

The effective theory has to reproduce the anomaly
relation~(\ref{QCD-anom}).  In the effective theory $\d_\mu j_\mu^\A$
can be found from the change of the action under the
transformations~(\ref{qU1A}) and (\ref{phiU1A}): $\delta
S=-\int\!d^4x\,\d^\mu\alpha\, j_\mu^\A =\int\!d^4x\,\alpha\,\d^\mu
j_\mu^\A$.  From the condition of anomaly matching one can deduce that
the effective Lagrangian contains the following terms,
\begin{equation}\label{Lanon-unexp}
\begin{split}
  \Lagrangian_{\rm anom} &= \frac1{8\pi^2 q_\A}
  \d_\mu\varphi_\A (e^2 C_{\A\gamma\gamma} A_\nu\tilde F^{\mu\nu} \\
   &\quad- 2e C_{\A\B\gamma}  B_\nu\tilde F^{\mu\nu}
   + C_{\A\B\B} B_\nu \tilde B^{\mu\nu})
\end{split}
\end{equation}
According to our previous discussion, each $B_\mu$ has to come with
$-\frac12\d_\mu\varphi_\B$.  Replacing $B_\mu\to
B_\mu-\frac12\d_\mu\varphi_\B$, and setting afterward $B_\mu=\mu
n_\mu$, we find
\begin{equation}\label{Lanon}
\begin{split}
  \Lagrangian_{\rm anom} &= 
  \frac1{8\pi^2q_\A} \d_\mu\varphi_\A
  \Bigl[e^2 C_{\A\gamma\gamma} A_\nu \tilde F^{\mu\nu}\\
  &\quad - 2 e C_{\A\B\gamma}
    \left(\mu n_\nu-{\textstyle\frac12}\d_\nu\varphi_\B\right)
    \tilde F^{\mu\nu}\\
  &\quad- {\textstyle\frac12}
  C_{\A\B\B}\, \epsilon^{\mu\nu\alpha\beta}
  \left(\mu n_\nu-{\textstyle\frac12}\d_\nu\varphi_\B\right)
  \d_\alpha\d_\beta\varphi_\B \Bigr]
\end{split}
\end{equation}
 
The term proportional to $C_{\A\gamma\gamma}$ in Eq.~(\ref{Lanon})
describes two-photon decays like $\pi^0\to2\gamma$ and
$\eta'\to2\gamma$.  Such processes occur already in the vacuum.  At
finite density $\pi^0\to2\gamma$ was considered previously in
Ref.~\cite{Nowak:2000wa} using a different technique, while
$\eta{'}\to2\gamma$ was considered in Ref.~\cite{Forbes:2001gj} using
the technique adopted in this paper. We do not consider these terms
here.  The new terms are the ones proportional to $C_{\A\B\gamma}$ and
$C_{\A\B\B}$.

At first sight, these new terms either vanish identically or are full
derivatives and cannot have any physical effect.  This is true when
the NG fields $\varphi_\A$ and $\varphi_\B$ are small fluctuations
from zero.  In particular, $\pi^0$ does not decay to a phonon and a
photon.  However, as $\varphi_{\A,\B}$ are periodic variables, the
action can be nonzero if either or both variables make a full $2\pi$
rotation.  This occurs in the presence of topological defects like
vortices or domain walls.

{\em 2SC and CFL phases}.---%
Our discussion so far has been rather general.  We now specialize
ourselves on two specific color superconducting phases: the two-flavor
color superconducting (2SC) and the CFL phases, and discuss the
topological defects in these phases.

In the 2SC phase, baryon number is not spontaneously broken, hence
the field $\varphi_\B$ does not exist.  
On the other hand, the U(1)$_\A$ symmetry is broken by a field
$\Sigma$ constructed from diquark condensates.
There is a single U(1)$_\A$ NG boson $\eta$ which corresponds to the
U(1)$_\A$ current $j^\A_\mu=\frac12\bar\psi\gamma^\mu\gamma^5\psi$
($Q^5_u=Q^5_d=1$).  $\Sigma$ has charge 2 under U(1)$_\A$: $q_\A=2$.
If the $\eta$ mass is sufficiently small, there are metastable domain
walls where $\varphi_\A$ changes by $2\pi$~\cite{SSZ-domain}.  The
boundary of such walls are closed $\eta$ vortices.  The coefficients
$C$'s are
\begin{equation}
  C_{\A\gamma\gamma} = \frac53\,,\quad
  C_{\A\B\gamma} = \frac13\,,\quad
  C_{\A\B\B} = \frac23
\end{equation}

In the CFL phase, the gauge-invariant order parameters are also
constructed from diquarks,
\begin{subequations}
\begin{eqnarray}
  X^{ai} &=& \epsilon^{ijk}\epsilon^{abc}\epsilon^{\alpha\beta}
        (q^{jb}_\alpha q^{kc}_\beta)^* \\
  Y^{ai} &=& \epsilon^{ijk}\epsilon^{abc}
  \epsilon^{\dot\alpha\dot\beta}
        (q^{jb}_{\dot\alpha} q^{kc}_{\dot\beta})^*
\end{eqnarray}
\end{subequations}
as $\Sigma= X^\+ Y$ which is a $3\times3$ matrix which breaks chiral
and U(1)$_\A$ symmetries.  One can also construct, e.g., $W = \det X\,
\mathrm{Tr}\, (X^{-1}Y)$ which breaks the baryon number.  In terms of
quark fields, $\Sigma\sim \bar q_{\rm L}^2 q_{\rm R}^2$, and $W\sim
\bar q_{\rm L}^4\bar q_{\rm R}^2$.

Obviously, the CFL phase has baryon vortices, see
Ref.~\cite{Forbes:2001gj} for details.  In fact, a rotating CFL core
of a star has to be threaded by a vortex lattice.  There are also
U(1)$_\A$ domain walls~\cite{SSZ-domain}.  The wall is actually not a
pure U(1)$_\A$ wall due to the mixing of $\eta$ and $\eta'$.  If one
parametrizes the three neutral axial NG bosons by
$\Sigma=\mathrm{diag}\,(e^{i\varphi_1},\, e^{i\varphi_2},\,
e^{i\varphi_3})$ then $\varphi_3$ is the lightest boson.  It is
generated by the current
\begin{equation}
  j^\A_\mu = \frac12\left(\bar u\gamma_\mu\gamma^5 u
  +\bar d\gamma_\mu\gamma^5 d
  -\bar s\gamma_\mu\gamma^5 s\right),
\end{equation}
i.e., $Q^5_u=Q^5_d=-Q^5_s=1$.  The transformation generated by
$j^\A_\mu$ leaves $\varphi_1$ and $\varphi_2$ unchanged, and changes
only $\varphi_3$, with $q_\A=2$.  The coefficients $C$'s are
\begin{equation}
  C_{\A\gamma\gamma} = \frac43\,,\quad
  C_{\A\B\gamma} = \frac23\,,\quad
  C_{\A\B\B} = \frac13
\end{equation}

In the CFL phase, the photon is mixed with a component of the gluon
field.  For colorless objects the effect of this mixing is small if
the electromagnetic coupling is much smaller than the strong coupling.
We will neglect this effect in the future.

{\em Nuclear matter}---%
The anomalous Lagrangian in nuclear matter can be derived in a similar
fashion.  We assume that neutrons  as well as protons form  
  a superfluid which
is characterized by the phases $\varphi_n$ and $\varphi_p$
correspondingly.
 We also allow for the neutrons
and the protons to have different chemical potentials, $\mu_n$ and
$\mu_p$ respectively.  The NG boson we consider is the $\pi^0$,
which is parametrized by the phase $\varphi_\pi$.  The axial current
is normalized so that $q_\A=1$; so $\varphi_\pi$ is related to the
canonically normalized $\pi^0$ field as $\varphi_\pi=\pi^0/f_\pi$,
$f_\pi\approx93~\textrm{MeV}$ in the vacuum.  The following
Lagrangian is obtained from Eq.~(\ref{Lanon}),
\begin{equation}
\label{Lanon-nuc}
\begin{split}
  \Lagrangian_{\rm anom} &= \frac{\d_\mu\varphi_\pi}{8\pi^2}
  \Bigl[e^2 A_\nu \tilde F^{\mu\nu} 
   + e\epsilon^{\mu\nu\alpha\beta}
  A_{\nu}  \d_\alpha\d_\beta\varphi_p  \\
 &  - \epsilon^{\mu\nu\alpha\beta}
  \left(\mu_p n_\nu - {\textstyle\frac12}\d_\nu\varphi_p\right) 
  \d_\alpha\d_\beta\varphi_p   \\
  &- \epsilon^{\mu\nu\alpha\beta}
  \left(\mu_n n_\nu - {\textstyle\frac12}\d_\nu\varphi_n\right) 
  \d_\alpha\d_\beta\varphi_n
  \Bigr]
\end{split}
\end{equation}

Armed with the effective anomalous Lagrangians~(\ref{Lanon}) and
(\ref{Lanon-nuc}), we now can discuss physical consequences of quantum
anomalies in densed matter.  We do not try to exhaust all effects, but
we would like to mention a few representative ones.

{\em Axial current on a superfluid vortex}.---%
To start, let us consider a superfluid (baryon) vortex.  For
definiteness, we assume the CFL phase, but many details are applicable
for nuclear matter as well.  Let the vortex be parallel to the $z$
axis and located at $x=y=0$.  Around the vortex $\varphi_\B$ changes
by $2\pi$, and in the center of the vortex it is ill-defined.  Outside
the core $\epsilon^{\mu\nu\alpha\beta}\d_\alpha\d_\beta\varphi_\B=0$,
and the term in $\Lagrangian_{\rm anon}$ that contains $\varphi_\B$
vanishes.  However, at the vortex core, as usual,
$(\d_x\d_y-\d_y\d_x)\varphi_\B=2\pi\delta^2(x_\perp)$.  The action
becomes
\begin{equation}
\label{action}
  S_{\rm anom} = \frac\mu{12\pi} \int\! dt\, dz\, \d_z\varphi_\A
\end{equation}
where the integral is a linear integral along the vortex line.  Now
let us recall that the axial current $J^\A_\mu$ is obtained, from
Noether's theorem, by differentiating the action with respect to
$\d_\mu\varphi_\pi$.  One sees immediately that {\em there is an axial
current running on the superfluid vortex}, with a magnitude of
$\mu/(12\pi)$.  This current, naturally, is coupled to the $Z$ boson.


From Eq.~(\ref{Lanon-nuc}) one finds an axial current on a neutron
superfluid vortex in nuclear matter.  This current can be understood
as a nonzero density of nucleon spin on the vortex.


{\em Axial boson--photon coupling on a baryon vortex}.---%
In the CFL phase there is the following term in the effective
Lagrangian,
\begin{equation}
  \d_\mu\varphi_\A \d_\nu\varphi_\B \tilde F^{\mu\nu}
  \sim
  \varphi_\A \d_\mu\d_\nu\varphi_\B \tilde F^{\mu\nu}
\end{equation}
This term is zero for topologically trivial field configurations but
does not vanish in the presence of vortices.  For a baryon vortex
located at $x=y=0$ we again set
$(\d_x\d_y-\d_y\d_x)\varphi_\B=2\pi\delta^2(x_\perp)$, and the
Lagrangian term becomes $\delta^2(x_\perp) \varphi_\A F_{03}$.  We
thus find that there is a linear coupling of neutral NG bosons with
the electric field in the vortex core.  Therefore, NG boson ($\pi^0$,
$\eta$, $\eta'$) striking the vortex line can be converted to one
photon and vibrations of the vortex line.  In the absense of vortices
these bosons can only decay into two photons.

Similar conclusion can be reached for $\pi^0$ striking the  proton superfluid
vortex in nuclear matter.

{\em Magnetization of axial domain walls}.---%
Let us consider an axial domain wall in an external magnetic field.
Such domain walls exist at very high densities where instanton effects
are suppressed~\cite{SSZ-domain}.  When the baryon field $B_\nu$ is
treated as a background, $B_\nu=(\mu,\vec 0)$, the following term is
present in the anomaly Lagrangian:
\begin{equation}\label{Bgrad}
  \Lagrangian_{\A\B\gamma} = \frac{eC_{\A\B\gamma}\mu}
  {8\pi^2}
  \vec{B}\cdot \vec{\nabla}\varphi_\A
\end{equation}
where we have set $q_\A=2$ and $\vec B$ is the magnetic field.
Consider now a U(1)$_\A$ domain wall streched along the $xy$
directions.  On the wall $\varphi_\A$ has a jump by $2\pi$.  Now turn
on a magnetic field perpendicular to the wall, i.e., along the $z$
direction.  Equation~(\ref{Bgrad}) implies that the energy is changed
by a quantity proportional to $BS$, where $S$ is the area of the
domain wall.  This means that {\em the domain wall is magnetized},
with a finite magnetic moment per unit area equal to
$eC_{\A\B\gamma}\mu/(4\pi)$.  The magnetic moment is directed
perpendicularly to the domain wall.  For the 2SC and CFL phases the
magnetic moment per unit area is $e\mu/(12\pi)$ and $e\mu/(6\pi)$,
respectively.

{\em Currents on axial vortices}.---%
The same effect can be looked at from a different perspective.  One
rewrites Eq.~(\ref{Bgrad}) into the following form,
\begin{equation}
  \Lagrangian_{\rm anom} = \frac{eC_{\A\B\gamma}\mu}{4\pi^2}\,
  \epsilon_{ijk} A_i\, \d_j\d_k \varphi_\A
\end{equation}
Since $\epsilon_{ijk}\d_j\d_k \varphi_\A\sim 2\pi\delta^2(x_{\perp})$
on the vortex core, the action can be written as a line integral along
the vortex,
\begin{equation}
   S_{\rm anom} = \frac{eC_{\A\B\gamma}\mu}{2\pi} 
   \int\!d\vec\ell\cdot \vec A
\end{equation}
which means that {\em there is an electric current running along the
core of the axial vortex}, which is similar to the axial current on
baryon vortices.  The magnitude of the current is given by
\begin{equation}\label{j}
  j^{\rm em} = \frac{eC_{\A\B\gamma}\mu}{2\pi}
\end{equation}
Taking $\mu\sim1.5~\textrm{GeV}$, in the CFL phase this is about
$40~\textrm{kA}$.


Naturally, the electromagnetic current running along a closed vortex
loop generates a magnetic moment equal to $\frac12 jS$, where $S$ is
the area of the surface enclosed by the loop.  This is exactly what we
found in Eq.~(\ref{Bgrad}).  A large vortex loop, therefore, has a
magnetic moment that can be interpreted as created by the current
running along the loop, {\em or} as the total magnetization of the
domain wall stretched on the loop.  The two pictures come from the
same term in the anomaly Lagrangian.

An axial vortex loop also carries angular momentum.  To see that, let
us place a vortex loop inside a rotating quark matter.  We thus put in
Eq.~(\ref{Lanon-unexp}) $B_\nu=\mu v_\nu$ where $v_\nu$ is the local
velocity and $\d_i v_j-\d_j v_i=2\epsilon_{ijk}\omega_k$.  The ABB
term in Eq.~(\ref{Lanon-unexp}) reads
\begin{equation}
  \Lagrangian_{\A\B\B} = \frac{C_{\A\B\B}\mu^2}{8\pi^2}
  \vec\omega\cdot\vec\nabla\varphi_\A
\end{equation}
which implies that the domain wall is characterized by a constant
angular momentum per unit area, equal to $C_{\A\B\B}\mu/(4\pi)$.  As
in the case of the magnetic moment, this angular momentum can
alternatively be interpreted as being carried by an energy (mass)
current running along the vortex surrounding the wall.

As is well known, for a relativistic system in order to have a nonzero
value for the angular momentum $ M^{ij} = \int\! d^3 x\, (x^i T^{0j} -
x^j T^{0i})$ one needs a time dependence such that $T^{0 i}\neq 0$.
Such time dependece is automatically appears if a system put into a
medium~\cite{magnus}.
In our case the construction of Ref.~\cite{magnus} is automatically
realized due to a nonzero value of $\mu$ in our system.



{\em Conclusions}.---%
We have seen that the quantum anomalies, especially those involving
the baryon current, lead to new and extremely unusual effects in
high-density matter.  The effects appear when topological objects
(vortices, domain walls) are present.  Some effects may have
consequences for the physics of compact objects, which are to be
explored. Of particular interest is the finding that baryon vortices
in the CFL phase and in nuclear matter carry weak neutral current,
which interact with neutrinos and may affect the cooling process of 
neutron stars.


Some of the effects discussed in this paper have precursors previously
discussed in the literature.  The currents flowing on vortices are
reminiscent of Witten's superconducting strings~\cite{Witten} and of
zero fermion modes on a vortex~\cite{JackiwRossi}.
Previously, Witten's construction
has been considered for dense 
matter~\cite{Kaplan:2001hh,Buckley:2002ur,Buckley:2002mx}.  In our
case, the currents on vortices appear in a much more direct fashion.

It should be said that there are many more physical effects of
anomalies that have not been considered in this paper.  They be
retrieved quite easily from the Lagrangian terms derived here.  One
particularly interesting effect is that there is an electric charge
located at the junction of a U(1)$_\A$ domain wall and a baryon
vortex, and that this charge is fractional.

All results in this paper have been derived from the anomaly terms of
the effective Lagrangian and do not rely on details of the microscopic
QCD Lagrangian, except for anomaly relations in the latter.  On the
other hand, it should be possible to understand the microscopic origin
of the vortex currents found here.  We plan to return to this question
in the future.

We have concentrated in this paper on a few phases of dense matter
(neutron superfluids, 2SC and CFL phases).  The results should be
extendable to other phases, e.g., phases with kaon and eta
condensation~\cite{Kryjevski:2004cw}.

The authors thank M.\,A.~Stephanov and D.\,B.~Kaplan for discussions.
A.R.Z. also thanks the Institute for Nuclear Theory at the University
of Washington for its hospitality and for organizing the workshop {\em
``QCD and Dense Matter: from Lattices to Stars''} when this work was
initiated.  The work of D.T.S. was supported, in part, by DOE grant
DE-FG02-00ER41132 and by the Alfred P.~Sloan Foundation. The work of
A.R.Z. was supported, in part, by the Natural Sciences and Engineering
Research Council of Canada.


\begin{thebibliography}{50}
\bibitem{Rajagopal:2000wf}
K.~Rajagopal and F.~Wilczek,
hep-ph/0011333.

\bibitem{Nowak:2000wa}
M.\,A.~Nowak, M.~Rho, A.~Wirzba, and I.~Zahed,
Phys.\ Lett.\ B {\bf 497}, 85 (2001).

\bibitem{Forbes:2001gj}
M.\,M.~Forbes and A.\,R.~Zhitnitsky,
Phys.\ Rev.\ D {\bf 65}, 085009 (2002)

\bibitem{SSZ-domain}
D.\,T.~Son, M.\,A.~Stephanov, and A.\,R.~Zhitnitsky,
Phys.\ Rev.\ Lett.\  {\bf 86}, 3955 (2001).

\bibitem{Kogut:1999iv}
J.\,B.~Kogut, M.\,A.~Stephanov, and D.~Toublan, 
Phys.\ Lett.\ B {\bf 464}, 183 (1999).

\bibitem{Kogut:2000ek}
J.\,B. Kogut, M.\,A. Stephanov, D.~Toublan, J.\,J.\,M. Verbaarschot, and
A.~Zhitnitsky, 
Nucl.\ Phys.\ B {\bf 582}, 477 (2000).

\bibitem{Son-superfluid}
D.\,T.~Son,
hep-ph/0204199.




\bibitem{magnus}
R.\,L.~Davis and E.\,P.\,S.~Shellard,
Phys.\ Rev.\ Lett.\ {\bf 63}, 2021 (1989).


\bibitem{Witten}
E.~Witten, 
Nucl.\ Phys.\ B {\bf 249}, 557 (1985).


\bibitem{JackiwRossi}
R.~Jackiw and P.~Rossi,
Nucl.\ Phys.\ B {\bf 190}, 681 (1981).




\bibitem{Kaplan:2001hh}
D.\,B.~Kaplan and S.~Reddy,
Phys.\ Rev.\ Lett.\  {\bf 88}, 132302 (2002).

\bibitem{Buckley:2002ur}
K.\,B.\,W.~Buckley and A.\,R.~Zhitnitsky,
J.\ High Energy Phys.\ {\bf 0208}, 013 (2002).

\bibitem{Buckley:2002mx}
K.\,B.\,W.~Buckley, M.\,A.~Metlitski, and A.\,R.~Zhitnitsky,
Phys.\ Rev.\ D {\bf 68}, 105006 (2003).

\bibitem{Kryjevski:2004cw}
See, e.g., A.~Kryjevski, D.\,B.~Kaplan, and T.~Sch\"afer,
hep-ph/0404290 and references therein.


\end{thebibliography}
\end{document}